# Spin wave dispersion and intensity correlation in width-modulated nanowire arrays: A Brillouin light scattering study

G. Gubbiotti,[1] L.L. Xiong,[2] F. Montoncello,[3] L. Giovannini,[3] and A.O. Adeyeye[2]

[1] Istituto Officina dei Materiali del CNR (CNR-IOM), Sede Secondaria di Perugia, c/o Dipartimento di Fisica e Geologia, Università di Perugia, I-06123 Perugia, Italy

[2] Information Storage Materials Laboratory, Department of Electrical and Computer Engineering, National University of Singapore, 117576 Singapore.

[3] Dipartimento di Fisica e Scienze della Terra, Università di Ferrara, Via G. Saragat 1, I-44122 Ferrara, Italy.

ABSTRACT

Using Brillouin light scattering spectroscopy and dynamical matrix method calculations, we study collective spin waves in dense arrays of periodically double-side width-modulated Permalloy nanowires. Width modulation is achieved by creating a sequence of triangular notches on the two parallel nanowire sides, with a periodicity of $p$=1000 nm, and tunable relative displacement ($\Delta$) of the notches sequence on the two lateral sides. Both symmetric ($\Delta$=0) and asymmetric ($\Delta$= 250 and 500 nm) width-modulated nanowires were investigated. We have found that the detected modes have Bloch-type character and belong to a doublet deriving from the splitting of the modes characteristics of the nanowire with homogeneous width. Interestingly, the amplitude of the magnonic band, the frequency difference of the doublet as well as their relative scattering intensity can be efficiently controlled by increasing $\Delta$ rather than having single- or symmetric ($\Delta$=0) double-side width-modulation.

Corresponding author:

Gianluca Gubbiotti

Email: gubbiotti@iom.cnr.it

ORCID: 0000-0002-7006-0370

INTRODUCTION

The growing requirements of spin waves (SWs) in a wide variety of phenomena (including logic gates for computing, insulating spintronics, wide frequency-operating range, nonlinear data processing, nonreciprocal devices and miniaturized magnetic components from microscale to nanoscale, etc) have resulted in new features and advances in magnonics technology. [1,2,3,4,5,6] All these fields of application depend on the possibility of engineering the SW band structure and are based on the capability of nanostructuring periodic magnetic materials on the nanoscale. [7,8,9]

Over the past decade, nanowires (NWs) with periodic geometrical constrictions have demonstrated as a useful tool for controlling the transmission band gap characteristics of SWs. NWs with a different periodic profile of the width modulation including rectangular, [10,11,12] sinusoidal [13] and asymmetrically sawtooth-shaped notched, [14] have been investigated by micromagnetic simulations. Xiong et al. studied symmetric width-modulated NWs and found that the number of detected modes depends on the NW thickness. This study, however, was limited to detect magnetization dynamics at the center of the first Brillouin zone (BZ), i.e., at a wave vector $k = 0$, where magnetization dynamics represents a family of in-plane standing SWs with zero Bloch wave numbers. [15] Recently, we reported an experimental and numerical investigation of the dispersion of collective spin waves of Bloch-type in single-side width-modulated NW array for two different value of the modulation periodicity ($p$ =500 and 1000 nm) and discussed how this affects the SW properties.[16]

In this work, we perform a systematic investigation by Brillouin light scattering (BLS) spectroscopy of the collective SW dispersion in double-side width modulated NW arrays as a function of the relative displacement ($\Delta$=0, 250 and 500 nm) between the edge corrugation on the two lateral NW sides. The obtained results indicate that modes of Bloch's type appear in doublets, localized in the wide and narrow NW portions. The novelties of the present work is that, when compared to the case of single-side width-modulated NWs, [16] a variation of $\Delta$ is a more efficient parameter for tuning the magnonic band of the lowest frequency mode. Also, the intensity ratio between the lowest frequency peak doublet significantly depends on $\Delta$.

All the experimental findings have been satisfactorily interpreted by the dynamic matrix method (DMM), which allows us to reproduce the band structure, to reconstruct the spatial profiles of the modes and to compute the BLS intensity spectra.

SAMPLE FABRICATION, EXPERIMENT and THEORY

The Permalloy ($Ni_{80}Fe_{20}$, Py) NW arrays were fabricated over an area of 4 × 4mm$^2$ on top of a silicon substrate using deep ultraviolet lithography at of 193-nm exposure wavelength leading to resit NW arrays. This was followed by deposition of Cr (5nm) /Py(30nm) using electron beam deposition in a vacuum chamber with a base pressure of $2.8 \times 10^{-8}$ Torr and at rate of 0.2 A°/s. The Cr film is used as adhesion layer. NWs with periodic width-modulation are obtained by etching a periodic sequence of triangular intrusions (notches) having width and depth fixed at 500 nm and 90 nm, respectively, on the two parallel sides of the NWs. The period of the width-modulation is fixed $p$ =1000 nm while edge modulations are displaced along the two lateral (left and right) NW edges by $\Delta$= 0, 250 and 500 nm. The first case corresponds to symmetric width-modulation while the other two cases ($\Delta$= 250 nm and 500 nm) to asymmetric one. An array of NWs with homogeneous (NW$_{NM}$) width $w$=350 nm and $d$=120 nm is also fabricated and used as reference sample. The NW arrays have periodicity $a=w+d$= 470 nm, which results in the edge of the first BZ located at $\pi/a = 0.66 \times 10^7$ rad/m. Scanning electron microscope (SEM) images of the NW arrays with different width-modulation, together with the corresponding measured magneto-optical Kerr effect (MOKE) loops, are shown in Fig.1.

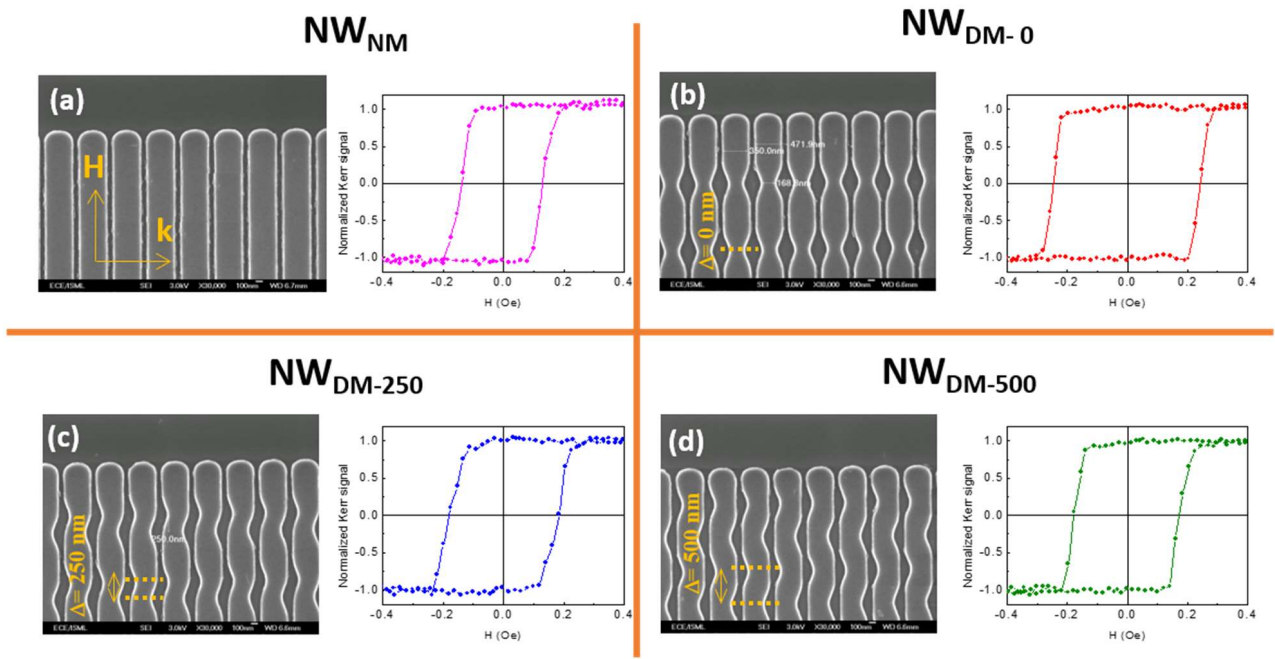

Fig. 1 SEM images of the investigated NWs arrays without (a) and with (b, c, d) double-side width-modulation together with the corresponding MOKE loops. The period of modulation ($p$) along the two sides of the NWs is fixed at $p$=1000 nm but the left modulation is shifted by $\Delta$= 0 nm, 250 nm and 500 nm with respect to the right one. Inset of panel (a) shows the direction of the applied magnetic field (H) and of the wave vector ($k$) while in panels (b-d) the displacement $\Delta$ is indicated by the arrow.

The MOKE hysteresis loops were measured at room temperature in the longitudinal configuration using a photoelastic modulator operating at 50 kHz and lock-in amplification. For all the investigated arrays, loops measured by sweeping the applied field along the NW length, are squared (easy direction) with a coercive field increment observed for the width $NW_{DM}$ when compared with $NW_{NM}$ ($H_c \approx 130$ Oe). This increment is more significant for the $NW_{DM}$ with $\Delta=0$nm ($H_c \approx 245$ Oe) and less pronounces for the $\Delta= 250$ nm and 500 nm NWs which have almost the same coercive field ($H_c \approx 180$ Oe). The $H_c$ enhancement in presence of width-modulation has been attributed to the DW pinning effect at the modulated regions.[17,18]

The BLS spectra were acquired in the backscattering configuration with the incident light, from a single-mode solid state laser operating at $\lambda = 532$ nm and with power of 200 mW, is focused over a sample area of about 40 micron in diameter by a camera objective of numerical aperture NA = 0.24 and focus distance of 50 mm. The frequency analysis is performed by using a (3+3)-tandem Fabry-Pérot interferometer with the incidence plane of light and the transferred wave vector ($k$) perpendicular to the applied magnetic field (Damon-Eshbach geometry). The SW dispersion was mapped for wave vector value ($k$) up to $2\times10^7$ rad/m, selected in the experiment, by properly choosing the appropriate incidence angle of light ($\theta$) according to the relation $k=(4\pi/\lambda)\times\sin(\theta)$. [19] A cross-polarization analysis on the scattered light has been performed to suppress the signal from elastically- and surface phonons-scattered light.

The SW frequencies, spatial profiles and BLS spectra were calculated by the DMM [20,21], which is based on the computation of a matrix, derived by the Hamilton equations of precession motion, and strictly dependent on the ground state configuration of the magnetization: for this matrix, an eigenvalue/eigenvector problem is solved at any wavevector value and direction. Eigenvalues and eigenvectors provide spin wave frequencies and spatial profiles, respectively. The sample is subdivided into a mesh of micromagnetic cells of $5\times5\times30$ nm$^3$, and the equilibrium magnetization configuration, used as input for the DMM dynamic calculations, is calculated by OOMMF micromagnetic code. [22] The dynamic magnetization of each mode can be interpreted using the expression: $\delta \boldsymbol{m_k} = \delta \widetilde{\boldsymbol{m}}_k e^{i\boldsymbol{k}\cdot\boldsymbol{r}}$ (Bloch wave), where $\delta \widetilde{\boldsymbol{m}}_k$ is the cell function, which has the periodicity of the array, and $\boldsymbol{k}$ is the wavevector in the Brillouin zone. In the following, we will plot the real z-component of $\delta \widetilde{\boldsymbol{m}}_k$ for $k=0$, which is mainly responsible for the BLS cross-section. [23] Standard material parameters for Py were used: saturation magnetization Ms=750 G, exchange stiffness constant A=$1.0\times10^{-6}$ erg/cm and gyromagnetic ratio ($\gamma/2\pi$) 2.95 GHz/$k$Oe.

Calculated modes are classified in terms of their spatial profiles as the quasi-uniform fundamental mode (F and F-loc) and $n$-DE and $n$-DE-loc (here DE stands for Damon-Eshbach) where $n$ is the number of nodal surfaces parallel to the NW length. For the F and $n$-DE modes the spin precession amplitude is mainly concentrated in the wide NW portion while the F-loc and $n$-De-loc it is localized in the narrow NW portion.

RESULTS and DISCUSSION

Fig. 2 presents the comparison between the measured and computed BLS intensity spectra at $k = 0$ (normal incidence of light upon the sample surface) for all the investigated NW arrays. The agreement is very good and the principal features of the spectra regarding either the the evolution of the relative peak intensity and frequency difference of the peak doublet are satisfactorily reproduced for different $\Delta$ values. For the $NW_{NM}$ (Fig. 2 (a)) only one peak is visible in the low-frequency range while a peak doublet is present for the $NW_{DM}$.

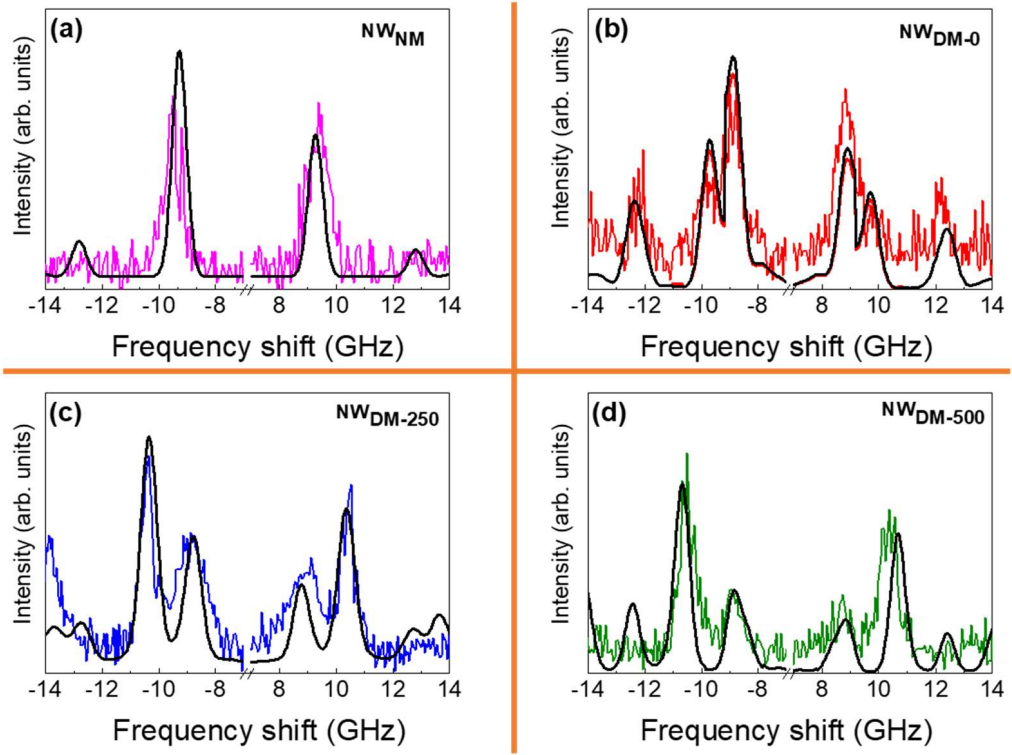

Fig. 2 Comparison between the measured (colored) and calculated (black) BLS intensity spectra recorded at $k=0$ and H= 500 Oe for all the NW arrays.

We found that there is a strong dependence of the peak intensity ratio between the lowest frequency modes on the $\Delta$ parameter for $NW_{DM}$. These two modes are identified as the F and F-loc modes, always being F at a lower frequency than F-loc. In particular, on increasing $\Delta$, the intensity ratio between those two dispersive modes (F and F-loc) reverses. For $\Delta=0$ (Fig. 2 (b)) the F mode is more intense than the F-loc, which appears as a weak shoulder at higher frequency. On the contrary, when $\Delta$ differs from zero (Fig. 2 (c) and (d)), the F-loc mode becomes more intense than the F one with an intensity asymmetry which increases passing from $\Delta=250$ nm to $\Delta=500$ nm. This behavior is related to the spatial extension of the mode which can be observed by a careful analysis of Fig. 4.

Fig. 3 reports the comparison between the measured and calculated dispersion curves (frequency vs $k$) for all the investigated NWs over the entire wave vectors accessible in our BLS setup. For the $NW_{NM}$ array only the lowest frequency mode exhibits a sizeable dispersion while higher frequency modes are dispersionless. This is different from what observed for the width-modulated NWs ($NW_{DM}$) where a doublet of peaks with sizeable dispersion is observed in the lowest frequency range of the spectra over the entire $k$-vector range investigated. An exception is for the $\Delta=0$ nm array, where F-loc mode is only visible close to the center of the Brillouin zone ($k=0$). Interestingly, the frequency difference between the two peaks of the doublet increases from 0.8 GHz for $\Delta=0$ nm up to about 1.6 GHz for both $\Delta=250$ and 500 nm. An overall good agreement between the measured and calculated dispersions is achieved. Slight disagreement is ascribed to the fact that calculations are performed by using the same set of magnetic parameters without the use of any adjustable value.

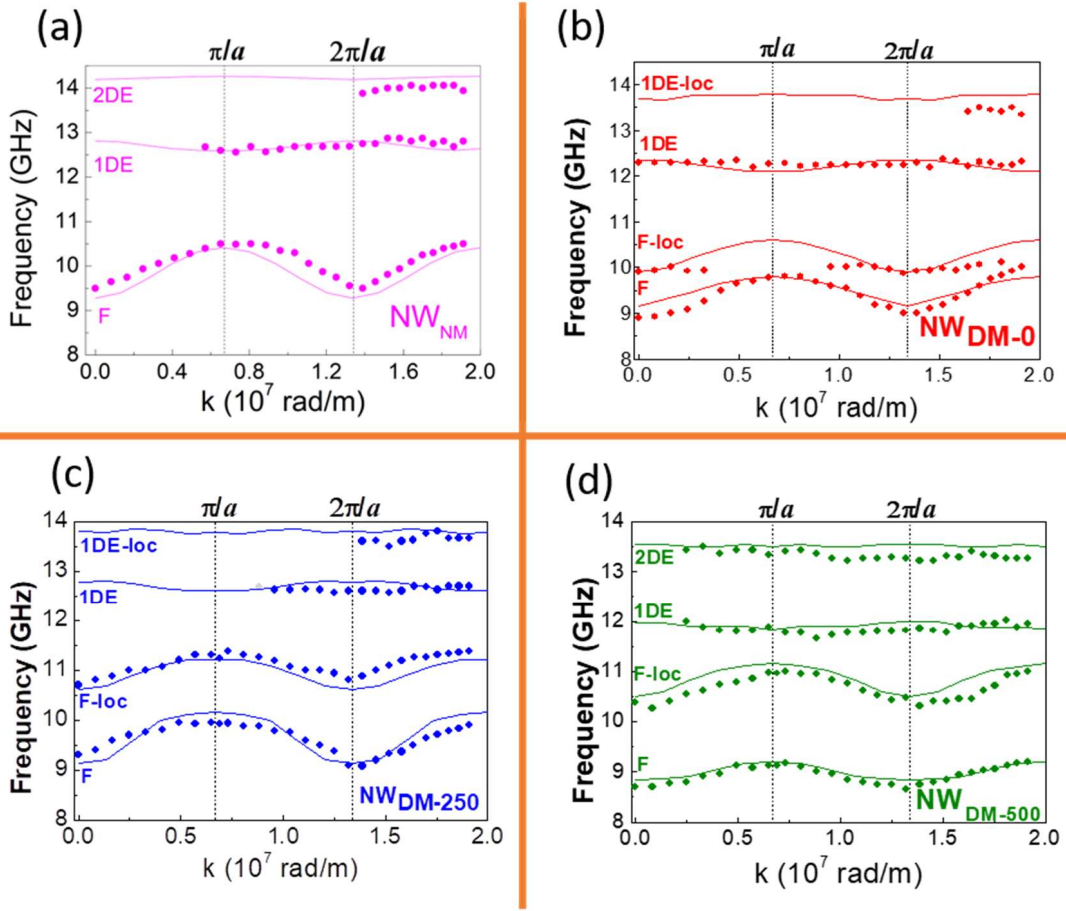

Fig. 3 Measured (points) and calculated (lines) frequency dispersions at H=500 Oe for (a) homogeneous NWs width ($NW_{NM}$) and NWs with vertical shift of the two lateral edges width-modulation of (b) $\Delta=0$nm, (c) $\Delta=250$ nm and (d) $\Delta=500$nm. Different labels of the calculated dispersion curves correspond to modes whose spatial profiles are reported in Fig. 4. The vertical dashed lines mark the boundary of the first BZ ($\pi/a$) and of the second BZ ($2\pi/a$).

To understand the observed dispersions, we calculate the spatial profile of SW for specific frequencies shown in Fig. 3. In Fig. 4, we plot the calculated out-of-plane component ($m_z$) of the dynamic magnetization for all the investigated NW arrays at $k=0$. We notice that in the $NW_{NM}$ and $NW_{DM-0}$ arrays, the mode profiles are symmetric with respect to the NW central axis while they are rather irregular for the asymmetric NWs ($\Delta=250$ and $500$ nm). This can be ascribed to the combined effects of the inhomogeneity of the internal field, described here below, and to the strong hybridization with other modes of backward character (having nodal plane perpendicular to the NW length).

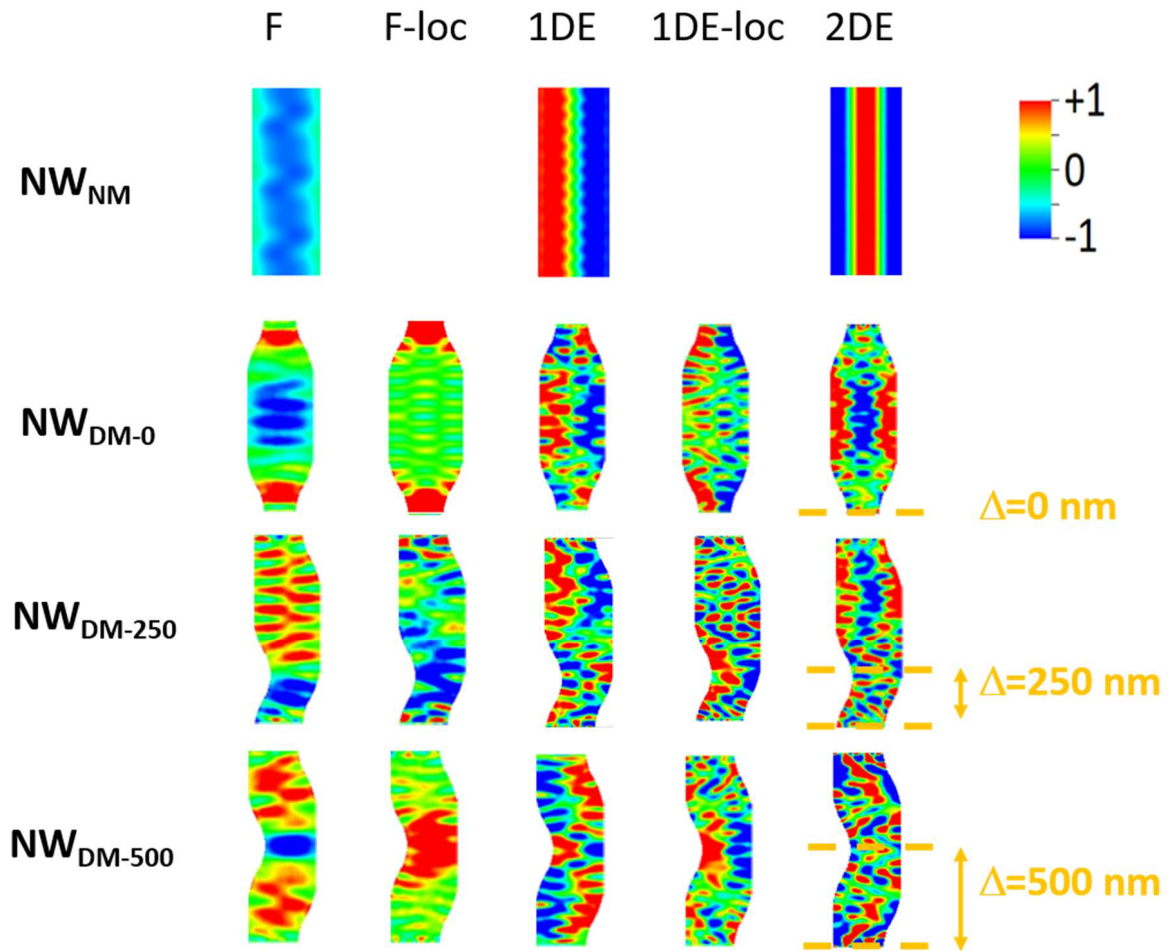

Fig. 4 Calculated out-of-plane component (real part) of the dynamic magnetization at $k=0$ of the principal modes detected by BLS spectra for $NW_{NM}$ and $NW_{DM}$ with different vertical displacement ($\Delta$) between the lateral sides corrugation. The magnetic field $H=500$ Oe is applied along the length of NWs (easy magnetization direction). Arrows indicate the relative displacement of the edge corrugation on the two NW lateral edges.

Interestingly, each mode observed for the $NW_{NM}$ splits into a doublet of modes localized in the wide and narrow portions of the $NW_{DM}$, whatever is the $\Delta$ value. Therefore, the F and F-loc modes observed for $NW_{DM}$ correspond to the quasi-uniform (F) mode of the $NW_{NM}$. Similarly, we have the 1DE ($1DE_{loc}$) and 2DE ($2DE_{loc}$) modes localized in the wider and narrower NW portions, respectively.

The origin of these additional modes found an explanation in terms of inhomogeneity of the effective magnetic field ($H_{eff}$) felt by the precessing magnetization. For the $NW_{NM}$, $H_{eff}$ is uniform with the magnetization parallel to the NW edges while the introduction of the width-modulation forces the magnetization to curl, following the lateral NW corrugation, thus deviating from the direction of the applied field. As a consequence, $H_{eff}$ is periodically modulated and the modes observed for the homogeneous width NW, split into doublets localized within regions with different $H_{eff}$ values and corresponding frequencies. This means that on increasing $\Delta$, the spatial regions with uniform magnetization are less extended and the F mode localization shrinks (blue areas in Fig. 4). This reflects on the fact that both its the frequency oscillation and peak intensity are strongly suppressed with respect to other arrays and to the F-loc mode (see spectra of Fig. 2). Regarding the amplitude of the magnonic band (amplitude of frequency oscillation) for the F mode, it linearly decreases on increasing $\Delta$ for $NW_{DM}$, passing from 0.9 GHz ($\Delta=0$), to 0.7 GHz ($\Delta=250$ nm) and finally to 0.4 GHz ($\Delta=500$ nm) while, for the F-loc, it increases from 0.6 GHz ($\Delta=250$ nm) to 0.8 GHz ($\Delta=500$ nm). The experimental value for the $\Delta=0$ nm NWs cannot be derived since the F-loc mode is not detected as $k$ approaches $\pi/a$. We have also found that the frequency difference between the F-Loc and the F mode increases linearly when $\Delta$ is increased from 0 to 500 nm. As a general comment, a magnonic band width of 0.8 GHz was previously measured for the F mode in single-side modulated NWs with periodicity of 1000 nm (see panel (b) of Fig.3 of Ref. [16]). In addition to this, a change of periodicity from 1000 nm to 500 nm for the single-side width-modulated NWs studied in Ref. 16, reduces the frequency width of the magnonic band for the F mode while leaves that of the F-loc mode almost unchanged to about 0.6 GHz. On the contrary, an increase of $\Delta$ for the double-side width-modulated NWs affects the magnonic band width of both the F and F-Loc modes even if with an opposite trend, as discussed above.

This contrasting behaviour is related to the fact that the bandwidth originates from the dynamic stray field generated by each mode through the array. This field has a dipolar nature and depends on the spatial extent of the regions where the modes are localized. Therefore, on increasing $\Delta$, the NW regions where the intrusions are present (absent) increase (decrease) and also the dipolar field generated by the F-Loc (F) mode increases (decreases). Our study also demonstrates that there is a direct correlation between the mode frequency bandwidth and the BLS intensity of the two lowest frequency modes since the BLS cross-section measured at $k=0$ is equal to the integral over the sample surface of the out-of-plane dynamic component of the magnetization. [23] These findings are very important as far as these two dispersive modes are considered for carrying and processing information in magnonic devices and suggest that the control of the intrusion offsets between the left and right NW sides is a more efficient way for tuning the magnonic band width rather than having single- or symmetric double-side width-modulation.

CONCLUSIONS

In conclusion, we experimentally demonstrate that the control over the dispersion of spin-wave in width modulated nanowires can be efficiently achieved through an asymmetric relative displacement of the periodic corrugations along the two nanowire lateral edges. In particular, we showed that this efficiency is related to the peculiar behavior of the main BLS active modes, which appear in doublets localized in different regions of the width-modulated nanowires and display different bandwidth and relative intensity depending on the relative displacement of the intrusions at the two nanowire sides. These results suggest a new way of tuning the magnonic band structure and can be used for the realization of magnonic devices enabling parallel data processing on multiple frequencies.

A.O.A. was supported by the National Research Foundation, Prime Minister's Office, Singapore under its Competitive Research Programme (CRP Award No. NRFCRP 10-2012-03).